# Study of electronic band alignment in SiGeSn/GeSn quantum well via internal photoemission effect

Justin Rudie,[1] Huong Tran,[2] Yang Zhang,[2] Sylvester Amoah,[2] Sudip Acharya,[2] Hryhorii Stanchu,[3] Mansour Mortazavi,[4] Timothy A. Morgan[5], Gregory T. Forcherio[5], Greg Sun[5], Gregory Salamo,[3,6] Wei Du,[2,3,*] and Shui-Qing Yu[2,3,*]

[1]Material Science & Engineering Program, University of Arkansas, Fayetteville, AR, 72701, USA
[2]Department of Electrical Engineering and Computer Science, University of Arkansas, Fayetteville, AR 72701, USA
[3]Institute for Nanoscience and Engineering, University of Arkansas, Fayetteville, AR, 72701, USA
[4]Division of Research, Innovation and Economic Development University of Arkansas at Pine Bluff, 1200 N. University Drive, 71601, USA
[5]Naval Surface Warfare Center, Electro-Optics Tech. Div. 300 Highway 361, Crane, IN 47522, USA
[6]Department of Physics, University of Arkansas, Fayetteville, AR 72701, USA
*Corresponding Author: weidu@uark.edu; syu@uark.edu

SiGeSn-based optoelectronic devices, which operate across a broad infrared wavelength range, have attracted significant attention, particularly heterostructures utilizing quantum wells are widely utilized. In these structures, band alignment type and barrier height are crucial for carrier confinement, making them highly desirable information to obtain. This work leverages the internal photoemission effect to extract effective barrier heights from a $Si_{0.024}Ge_{0.892}Sn_{0.084}$ / $Ge_{0.882}Sn_{0.118}$ single quantum well structure, which was pseudomorphically grown on $Ge_{0.9}Sn_{0.1}$ and Ge buffered Si substrate. The extracted effective barrier heights are approximately 22±2 and 50±2 meV for electrons and holes, respectively. Moreover, we have identified the type-I band alignment between GeSn well and SiGeSn barrier, as indicated by an internal photoemission threshold of 555±1 meV.

GROUP-IV materials such as GeSn/SiGeSn have garnished increasing attention due to their potential for efficient Si-based optoelectronics.[1-6] GeSn/SiGeSn technology could be a key enabler for low-cost and high-performance photonic integrated circuits on Si photonics platform by leveraging the mature complementary metal-oxide-semiconductor (CMOS) processes. The compelling factors of SiGeSn alloys include: i) true direct bandgap which is desirable for band-to-band lasers [7]; ii) all-group-IV materials allowing for SiGeSn-based devices to be monolithically integrated on Si substrate;[3] iii) tunable bandgap enabling broad wavelength coverage from near- to mid-infrared region [7]; iv) bandgap energy and lattice constant that can be independently engineered, making it possible to form desirable type-I band alignment, which is favorable for quantum confinement [8].

The band offset type and magnitude at a heterojunction interface are key parameters for designing heterostructure-based devices, broadly acknowledged as key building blocks in double heterostructures (DHS) and quantum wells (QW) devices employed for their favorable carrier confinement in lasers and photodetectors [9]. One effective method for measuring band offset is to employ internal photoemission effect (IPE). This process involves transfer of photoexcited carriers from one material to another when the incident photon energy exceeds the barrier height, resulting in a measurable barrier height that corresponds to the band offset in either conduction band (CB) or valence band (VB) [10, 11].

Band alignments have been the subject of intensive research on group IV, III-V and II-VI material systems to facilitate the heterostructure design [12-18]. Band offsets among alloys of SiGeSn containing different Si and Sn compositions have not been determined experimentally. Theoretical calculations often rely on model approximation with reasonable interpolations- it is therefore highly desirable to obtain such experimental information to unlock the potential of advanced electronic/photonic devices [19].

In this work, the band offset of a $Si_{0.024}Ge_{0.892}Sn_{0.084}$ / $Ge_{0.882}Sn_{0.118}$ single quantum well (SQW) was experimentally characterized via IPE. It is worth noting that in QW structures, the measured IPE threshold corresponds to the effective barrier height between the quantized energy level in the well and band minima in the barrier (maxima) in CB (VB), which is smaller than the absolute band offset in bulk material (see Fig.2 (a)). However, since in QW structures the effective carrier confinement is determined by the relative barrier height rather than the absolute band offset, making the results obtained in this study more useful to evaluate the carrier confinement for device design [20]. A type-I band alignment was identified for the SQW sample in this work with effective barrier heights of ~22 and 50 meV in CB and VB, respectively.

Figure 1(a) shows a conceptual schematic of electron IPE process. A heterojunction structure consists of a narrower bandgap material ($E_{g1}$) and a wider bandgap material ($E_{g2}$). As the photon energy of incident light reaches a certain value (IPE threshold), the photo-excited electrons could be transited from the VB of narrower bandgap material to the CB of wider bandgap material across the interface (labeled as ②), leading to the sharply increased photocurrent. In addition to IPE process, two other processes also contribute to photocurrent: 1) photo-enhanced thermionic emission labeled as ①, and 2) light absorption in wider bandgap material labeled as ③. Similarly, the hole IPE process is illustrated in Fig. 1(b). The photon energy that triggers process ② can be extracted as IPE threshold from the photocurrent curve (as a function of incident photon energy). Subsequently, the band offset can be obtained by subtracting the narrow bandgap energy from the IPE threshold [21]. Note that bias induced band bending at interface also affects the effective barrier height, which, however, is small relative to the band offset, and therefore it can be ignored in IPE analysis.

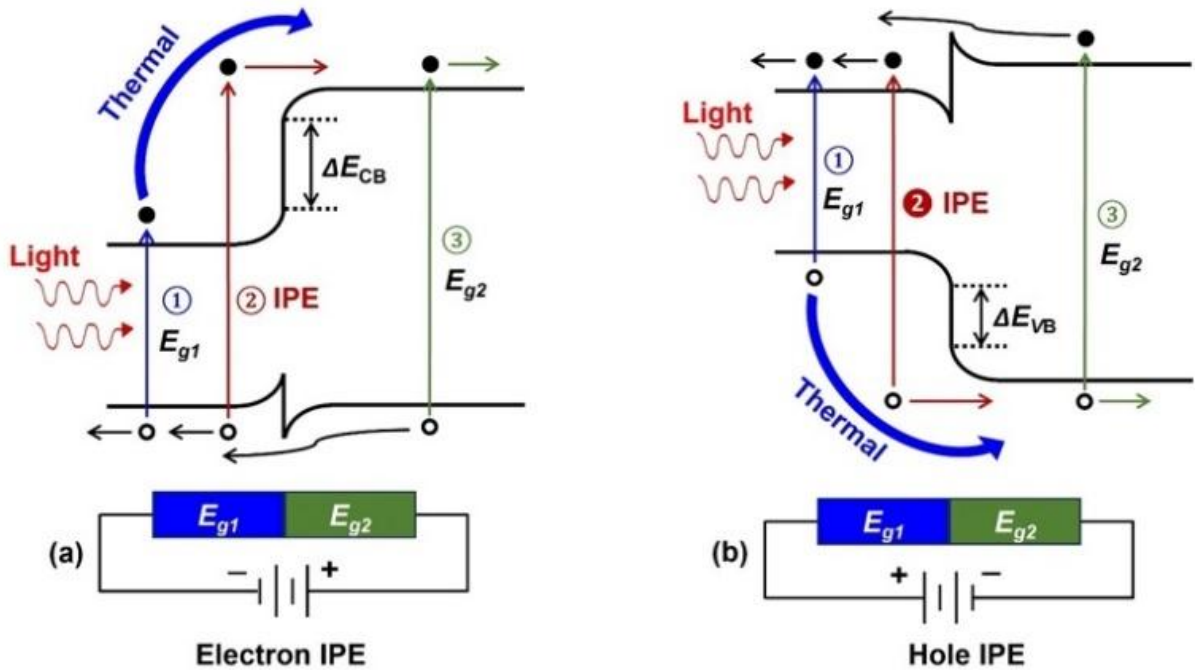

Fig. 1 Concept schematics of (a) electron IPE and (b) hole IPE. Processes ①, ②, and ③ corresponds to absorption in material 1 ($E_{g1}$), electron (hole) IPE, and absorption in material 2 ($E_{g2}$), respectively.

IPE effect can also be observed in QW, as studied in this work. Figure 2(a) shows the IPE effect in a SQW structure, assuming the type I band alignment. Process ① associated with the absorption in the well occurs at relatively lower photon energy between the ground quantization energy levels in CB and VB. If the band offsets are significantly different in the CB and VB, technically two thresholds can be observed in photocurrent curve as photon energy increases, corresponding to electron (labeled as ②) and hole (labeled as ❷) IPEs, respectively. On the other hand, these two IPE processes may not be identifiable if the two band offsets in CB and VB are close. As photon energy further increases, process ③ occurs as the direct absorption in the barrier. Each process would lead to a noticeable knee in the photocurrent curve, characterized by a sudden increase in the current as the photon energy surpasses each threshold.

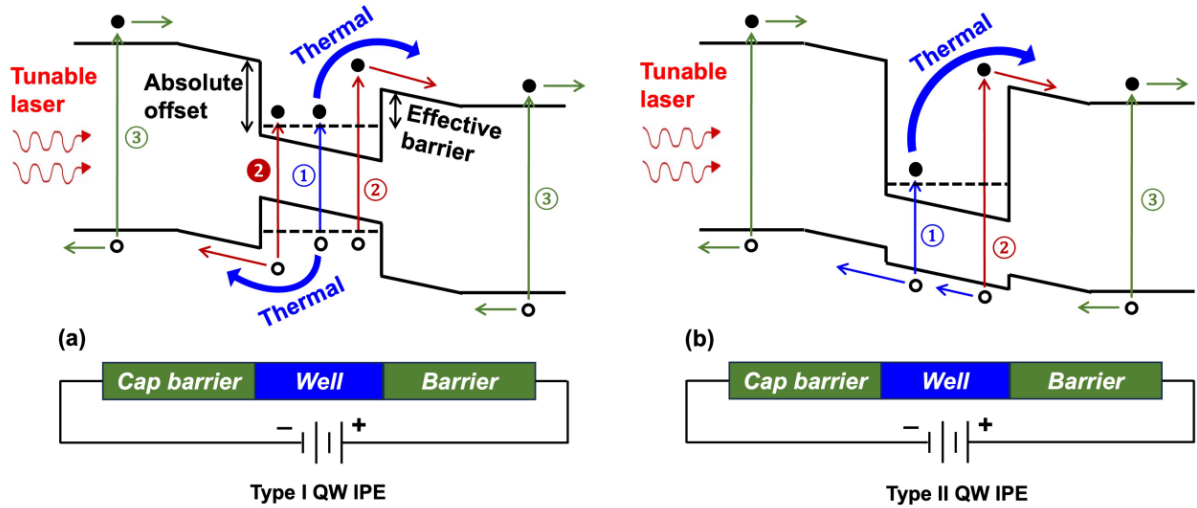

Fig. 2 Schematics of IPE effects in (a) type-I SQW and (b) type-II SQW. Processes ①, ②, and ③ corresponds to photon absorption in well, electron (hole) IPE, and absorption in barrier, respectively.

A SQW with type-II band alignment should show distinctly different photocurrent behavior compared to the type-I case. As illustrated in Fig. 2(b), the absence of a barrier for holes in the well results in only electron IPE taking place. In this scenario, the excitation energy must exceed the bandgap energy of the barrier, causing the absorption in the barrier (process ③) to occur prior to the IPE process. Consequently, the IPE threshold should be observed following the barrier absorption, provided it can be differentiated from process ③.

In this work, the $Si_{0.024}Ge_{0.892}Sn_{0.084}$ / $Ge_{0.882}Sn_{0.118}$ SQW layer stack was epitaxially grown using an industrial standard ASM Epsilon® 2000 Plus reduced pressure chemical vapor deposition (RPCVD) reactor. Commercially available $SiH_4$, $GeH_4$, and $SnCl_4$ were used as precursors. A ~1000-nm-thick Ge layer was grown on the Si substrate using a two-step method, followed by the GeSn buffer growth with gradient Sn compositions from 6% to 9% to ease the compressive strain in the GeSn well. The SQW layer stack consists of a 39-nm-thick $Ge_{0.882}Sn_{0.118}$ well sandwiched between two 72-nm-thick $Si_{0.024}Ge_{0.892}Sn_{0.084}$ barriers (Fig. 3(a)). The detailed material growth can be found elsewhere [22, 23].

The sample was fabricated into square mesa devices with side lengths of 500 μm using standard photolithography technique, followed by a wet chemical etching process using the mixture of HCl, $H_2O_2$, and DI $H_2O$ at a ratio of 1:1:10 to form the mesa. An ice & water bath was used to control the etching temperature at 0 °C. Both top and bottom contacts were formed by depositing 10/300 nm Cr/Au on the SiGeSn cap and SiGeSn bottom barrier layers. Our previous work confirmed that the metal-semiconductor interface is Ohmic contact [24, 25]. The measurement setup consists of a tunable laser operating at 1.9-3.0 μm (0.41-0.65 eV) as the excitation source, a lock-in amplifier to collect the chopped signals from a known resistance, and a source measure unit to provide bias voltage. Devices under test are wire-bonded in a cryostat for temperature-dependent characterization.

The SQW sample was analyzed using secondary ion mass spectrometry (SIMS) for its elemental compositions and layer thicknesses, as shown in Fig. 3(a). The results are consistent with the designed parameters, i.e., 2.5% Si and 8.3% Sn in the SiGeSn barrier and 11.8% Sn in the GeSn well. Although the bottom SiGeSn is slightly thicker than the top layer, it should not affect the IPE characteristics. Band diagram calculations indicate that the bandgap energy difference between the GeSn well and SiGeSn barrier is sufficiently large (~80 meV) for the IPE effect to be identified. Figure 3(b) shows the x-ray diffraction (XRD) rocking curve. The black curve represents the measured data, while the red curve denotes the simulation data. The Ge buffer peak is located at 66°, consistent with results reported elsewhere [26]. A broad peak at ~65° consists of combined contributions from the GeSn buffer and SiGeSn barrier, which overlap with each other and therefore cannot be further identified. GeSn QW peak appears at ~64°, agreeing well with theoretical studies [27]. The clear resolution of each peak indicates high material quality. In addition, the XRD simulation was performed using X'Pert Epitaxy software associated with XRD instrument. By specifying the degree of strain relaxation of each layer using the data obtained from reciprocal space map (RSM, see Fig. 3(c)), the lattice constant can be calculated, and then the Sn composition can be calculated using the Vegard's law (linear interpolation of between Ge and Sn). The Sn compositions extracted from simulation match well with the SIMS data. The Si and Sn compositions were determined by the cross check of SIMS and XRD characterization results.

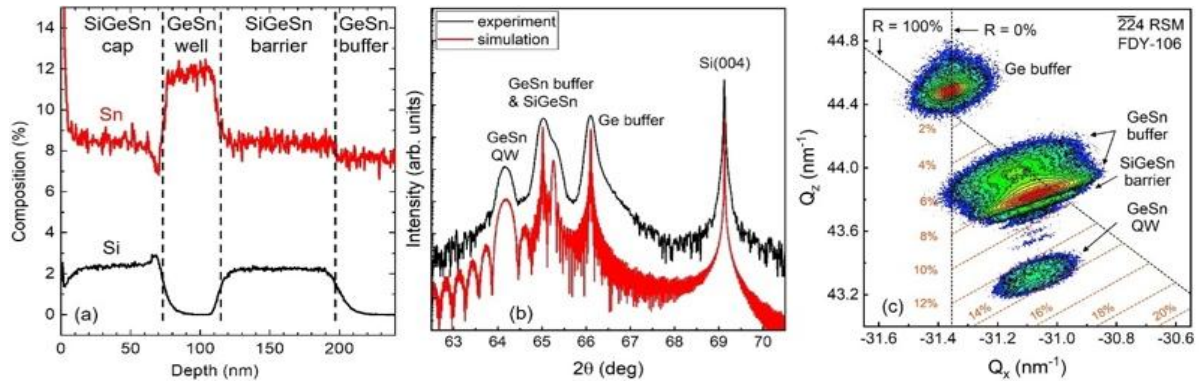

Fig. 3 (a) SIMS showing Si and Sn compositions and layer thickness. (b) XRD rocking curve. Black and red curves are measured data and simulation, respectively. (c) RSM contour plot showing the strain of each layer. The percentage indicates the Sn composition. The R=0% and 100% lines indicate the full strain and full relaxation, respectively.

The reciprocal space map (RSM) contour plot in Fig. 3(c) shows the strain of each layer. The GeSn buffer is almost fully relaxed with a residual compressive strain of ~0.8%. The broadened contour of GeSn results from the gradient Sn composition from 6% to 9%. The QW layer stack sitting above the GeSn buffer consisting of the SiGeSn barrier and the GeSn well is grown pseudomorphically, i.e., lattice matched to the GeSn buffer, and thus the GeSn well experiences in-plane compressive strain of 8.2% and out-of-plane tensile strain of 6.2%, respectively. The RSM data confirms the Sn compositions in each layer.

Photoluminescence (PL) spectroscopy was performed using a standard off-axis configuration with a lock-in amplifier. A 532 nm continuous wave (CW) laser (100 mW) and a 1064 nm nano-second pulsed laser (45 kHz repetition rate and 2 ns duration, 100 mW average power) were used as the excitation source. The PL emission was collected by a spectrometer and then fed into an InSb detector with a cutoff wavelength of 5.0 μm.

Figure 4(a) shows the PL spectra under 532 nm laser excitation. Due to the shallow penetration depth of less than 50 nm (see Fig. 4(a) inset), the majority of light absorption occurs in the top SiGeSn barrier, the excited carriers subsequently diffuse into the GeSn well, the PL data is thus dominated by the carrier recombination in the well. The peak is attributed to the direct bandgap emission from the GeSn QW. The PL intensities at lower temperatures appear to be significantly stronger, which is a typical characteristic for direct bandgap emission.

Figure 4(a) inset shows the Arrhenius plot derived from the temperature-dependent PL spectra. As temperature increases, the PL intensity decreases due to enhanced non-radiative recombination. The data were fitted using the standard thermal quenching model to extract the activation energy associated with non-radiative recombination, which is approximately 50.7 meV. It likely corresponds to a defect-related non-radiative channel. The activation energy is roughly $7.7\times k_BT$ at 77 K, indicating that the thermally activated escape of carriers is suppressed.

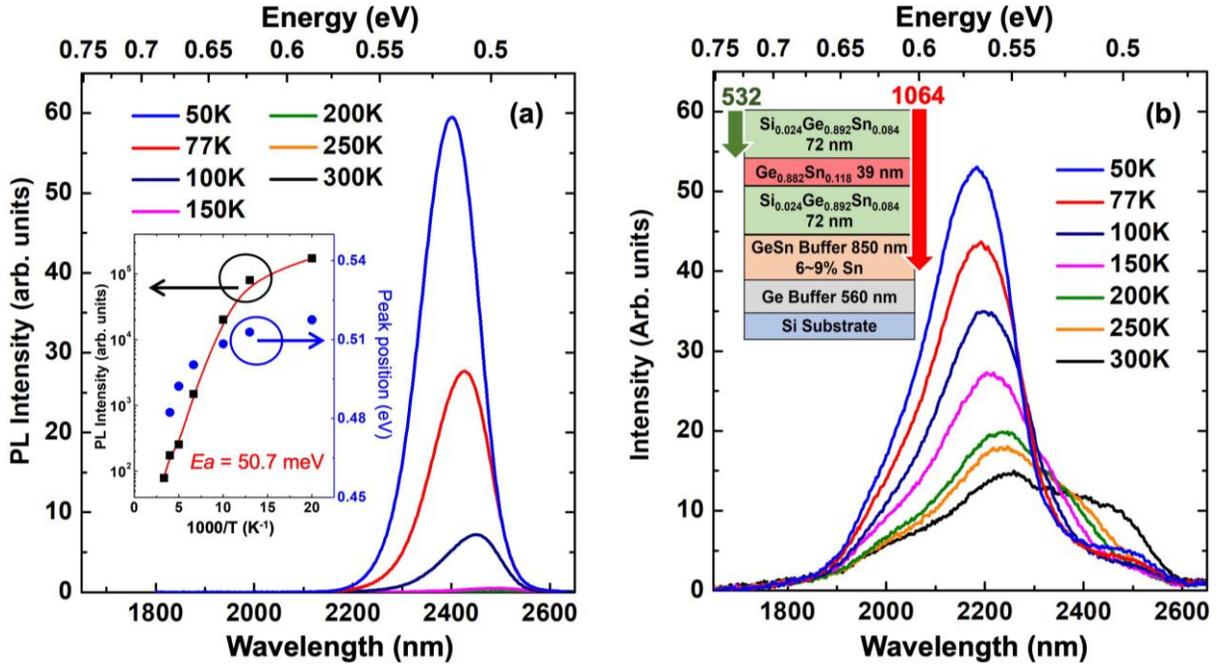

Fig. 4. Temperature-dependent PL spectra under (a) 532 nm CW laser and (b) 1064 nm pulsed laser pumping. Inset in (a): Arrhenius plot of integrated PL intensity as a function of inverse temperature (1000/T). The activation energy was extracted as 50.7 meV. The temperature-dependent emission energy was also plotted.

Figure 4(b) shows the temperature-dependent PL spectra under 1064 nm pulsed laser excitation. Due to the greater penetration depth, optical absorption occurs in all heterostructure stack layers, and thus the thicker SiGeSn barriers (top and bottom) have much higher light absorption than GeSn well. Considering the instantaneously high peak power and relatively short carrier lifetime, the photo-generated carriers in SiGeSn barriers could recombine before they are collected by GeSn well. As a result, the emission from the SiGeSn barrier dominates the PL spectra, even SiGeSn barrier is an indirect bandgap material, appearing around 2200 nm in Fig. 4(b). Additionally, a longer wavelength shoulder at ~2500 nm is attributed to the QW emission, consistent with the PL peak observed in Fig. 4(a).

The PL spectroscopy study provides optical transition energies in the well (between ground energy levels in CB and VB) and in barrier. The PL data is essential to facilitate the IPE analysis. Notably, the transition energy difference is ~72 meV at 77 K, which reflects the total effective barrier heights in CB and VB and is sufficient to differentiate the IPE threshold energy even with a relatively small band offset.

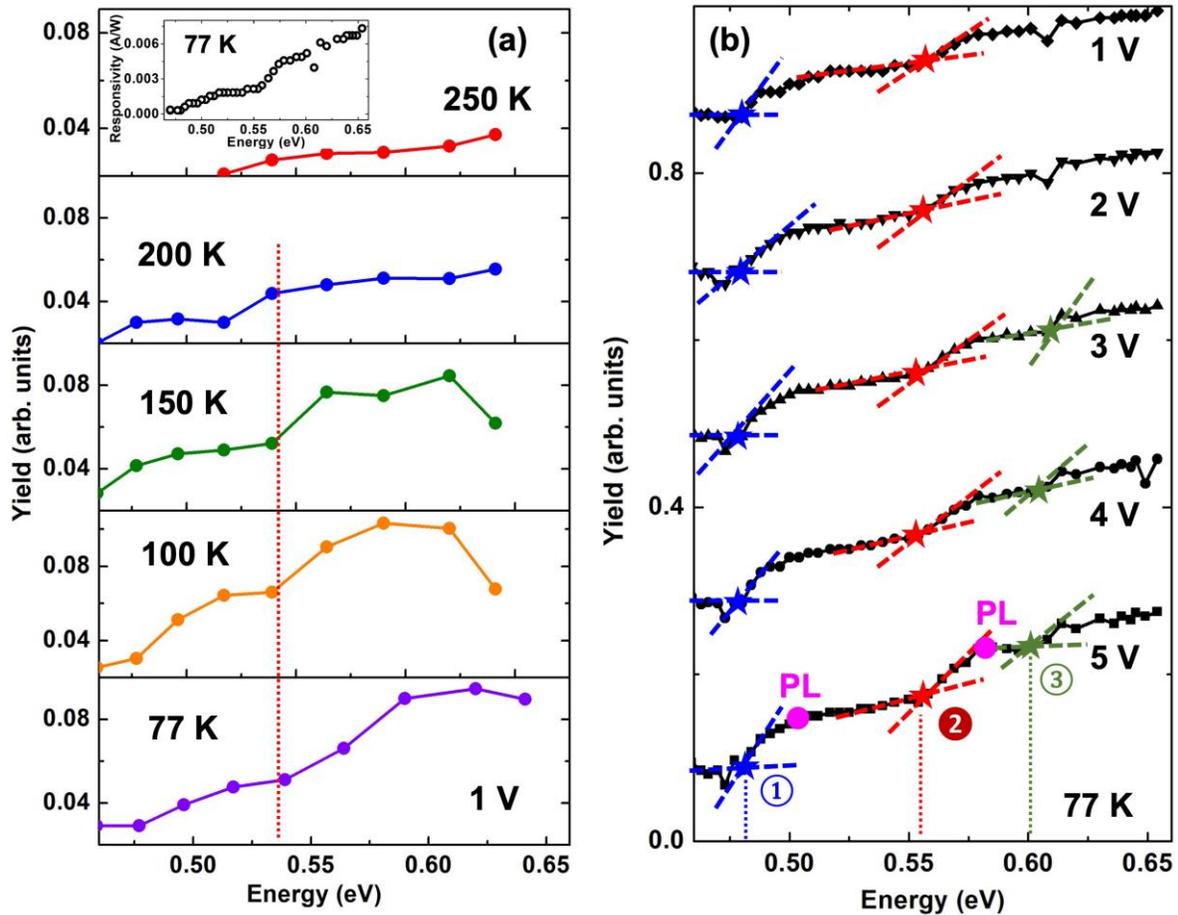

Fig. 5. (a) Temperature-dependent quantum yield at 1 V bias voltage. The dotted line is eye guidance. Inset: responsivity at 77 K. (b) High-resolution quantum yield at 77 K with various bias voltages. The energies associated with different absorption mechanisms were extracted by data fitting (dashed lines). Curves in (b) are stacked for clarity.

IPE is characterized by the quantum yield, which is defined as the number of emitted carriers per one incident photon. It is usually expressed as the product of responsivity and photon energy [18], [28]. The IPE quantum yield $Y$ can be extracted from photo current by normalization with respect to the incident photon flux and is expressed as a power function of the excess photon energy above the threshold energy: $Y(hv) = A(hv - \phi)^P$ [29], where $\phi$ corresponds to band-to-band absorption energy or effective barrier height at the interface, which is directly associated with the IPE process threshold. The exponent $P$ is 3 when electrons are excited out of the VB [30]. Generally, quantum yield curve exhibits a sharp increase when a "threshold" is reached, due to the dramatic enhancement in responsivity, followed by a more gradual increase until the next "threshold" is encountered as the photon energy $hv$ increases. Since thermally excited carriers could lower the signal-to-noise ratio of measured quantum yield curve, the low temperature quantum yield characterization would provide a better insight into the intrinsic band structure.

Figure 5(a) shows the normalized quantum yield at various temperatures under 1-V bias voltage. At temperatures above 200 K, the quantum yield gradually increases along with the photon energy, with no observable IPE threshold due to the background thermal excitation. Below 200 K, however, a clear threshold appeared at 0.54 eV in each curve even with low-resolution for quick scan, indicating the IPE process occurs at this photon energy. The dotted line serves as a guide for the eye. As temperature decreases, the threshold energy is almost unchanged, indicating that temperature has little or no influence on band offset.

To further study the IPE process, quantum yield at 77 K was characterized with high-resolution under bias voltages ranging from 1-5 V, as shown in Fig. 5(b). At each voltage, three distinct kinks can be clearly identified at 0.485, 0.555, and 0.601 eV. These values are extracted through data fitting as indicated by the dashed lines with the error bar of ±1 meV. At higher voltages, the threshold energies are more pronounced because of increased band bending. The kink at 0.485 eV (labeled as ①) originates from the band-to-band photon absorption in GeSn QW, as it is close to the measured PL peak from the GeSn well at 77 K shown in Fig. 4(a) (508±2 meV). Likewise, the kink at 0.601 eV (labeled as ③) is assigned to the band-to-band photon absorption in SiGeSn barrier, whose PL peak is at 580±2 meV.

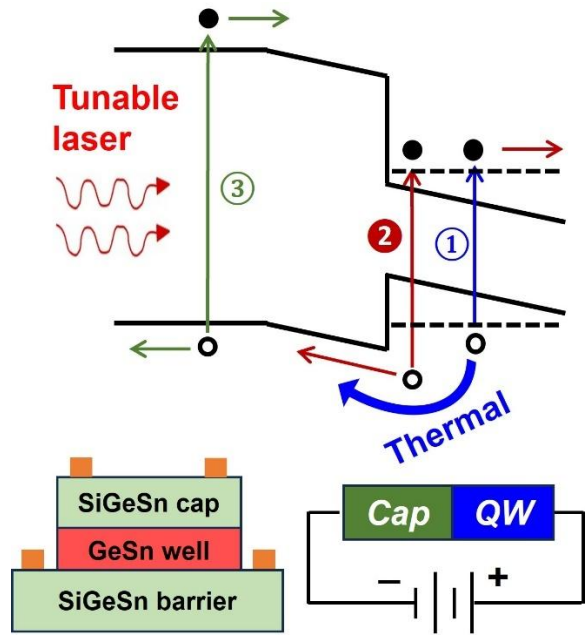

Fig. 6. Transitions in SiGeSn/GeSn QW. Hole IPE was observed in this structure. Bottom left: device schematic. Bottom right: applied bias voltage.

To identify the responsible IPE process, the band diagram was plotted in Fig. 6 according to device fabrication procedure. The kink at 0.555 eV (labeled as ②) corresponds to the hole IPE process as the holes flow from narrower bandgap GeSn well towards wider bandgap SiGeSn barrier/cap following the absorption of incident photons. It is worth noting that: i) the measured quantum yield starts to increase at 0.485 eV because of the distinctly enhanced photon absorption, which could occur at the photon energy below bandgap due to the Urbach tail, a band tail in an absorption spectrum arises primarily from carrier trapping in a structural, alloy, or phonon-related disorder

[31, 32]. Urbach tail energy ranges from a few meV to a few tens meV. Therefore, the energy difference of 23 meV between PL peak at 0.508 meV and ① in IPE of 0.485 meV might be due to the band tail absorption; ii) the measured PL peak of 0.580 meV is assigned to indirect bandgap emission from the barrier, while the ③ in IPE (0.601 eV) may correspond to the direct bandgap absorption. The absorption coefficient of indirect bandgap is relatively lower compared to that of direct bandgap, resulting in more pronounced increase of quantum yield at 0.601 eV. The SiGeSn barrier is an indirect bandgap material. Based on our previous bandgap calculations and the PL behavior shown in Fig. 4, the PL spectra are dominated by indirect transitions; and iii) the measured intensity of PL can be modeled as [33]:

$$I(h\nu) \propto (h\nu - E_{tr})^{1/2} \exp\left[-(h\nu - E_{tr})/kT\right]$$

where $h\nu$ is photon energy and $E_{tr}$ is optical transition energy. The peak of intensity can be found from the derivative of above equation, which gives:

$$E_{tr} = E_{PL} - (1/2)kT$$

Here, the $E_{PL}$ is the measured PL peak. For QW emission it is 508±2 meV. The transition energy $E_{tr}$ is energy difference between the ground quantization energy levels in CB and VB of GeSn QW. At 77 K, $E_{tr,QW} = 505 \pm 2\ meV$. Likewise, transition energy in barrier is $E_{tr,barrier} = 577 \pm 2\ meV$.

The effective barrier heights in VB and CB can then be extracted from the energy difference between ② and $E_{tr,QW}$, and between $E_{tr,barrier}$ and ②, which are 50±2 and 22±2 meV, respectively. These values indicate that the hole confinement is stronger than the electron confinement. Note that these effective barrier heights were estimated at 77 K. Considering the bandgap shrinkage at higher temperature, the effective barrier height may be reduced. For practical device applications, for example, lasers operating in room temperature, these barrier heights are considered insufficient compared to theoretical requirement of ~100 meV [32].

It is worth noting that the band alignment type in SiGeSn/GeSn material system has been explored only theoretically, with no direct measurement results reported prior to this study. The threshold characteristic shown in Fig. 5 clearly reveals the type-I band alignment for the SQW sample in this work. If the band alignment were to be type-II, either electron or hole IPE threshold would be higher than SiGeSn barrier absorption energy (process ③), and therefore no kink can be observed between processes ① and ③. The existence of 0.555 eV threshold between ① and ③ unambiguously confirms the type-I band alignment.

In addition to processes ①, ❷, and ③, the defect, donor, and acceptor dopants levels may be involved in optical transitions as well. However, normally their activation energies are significantly smaller than the starting energy of tunable laser (0.45 eV), and therefore they would act as the constant background noises to quantum yield measurement. As a result, they won't affect the estimation of effective barrier heights. Moreover, the excitonic effects were not considered in this work, since they are unknown in SiGeSn material system. More studies are needed to understand the excitonic effects in optical transitions.

IPE measurements were conducted on a $Si_{0.024}Ge_{0.892}Sn_{0.084}$ / $Ge_{0.882}Sn_{0.118}$ SQW sample to determine the effective barrier height and band alignment type. Sample quality and compositions were measured by SIMS and XRD. PL measurements provided the transition energies in the GeSn well and SiGeSn barrier, allowing to extract the barrier heights in the CB and VB as 22±2 and 50±2 meV, respectively. Furthermore, the hole IPE threshold of 0.555 eV unequivocally confirms a type-I band alignment for the SQW sample. We would like to mention that theoretically, there should be four types of band offset with strained quantum well structure: Γ-Γ valley and L-L valley in CB, and heavy hole (HH)-HH and light hole (LH)-LH band in VB. At this stage, it is very difficult to distinguish each type of band offset based solely on the measured quantum yield. However, the measured IPE threshold effectively reflects the carrier confinement, which is more relevant to device design than the absolute band offset. Therefore, the effective barrier height was used to represent the overall excitation process. Although the band offset is not directly measured, the Type I band alignment type can be unambiguously identified, and the carrier confinement can be reasonably estimated through this work.

The authors would like to acknowledge the support by DoD Multidisciplinary University Research Initiatives (MURI) (Grant No. FA9550-19-1-0341). J.R. acknowledges support from the DoD SMART Fellowship. G.T.F would like to acknowledge the support by the Naval Innovation Science and Engineering (NISE) program and Office of Naval Research (ONR) Code 312.

The authors have no conflicts to disclose.